\documentclass[twocolumn,floatfix,showpacs,preprintnumbers,superscriptaddress,amsmath,amssymb,aps,prl]{revtex4-1}
\usepackage{amssymb,amsmath,graphics,epsfig,amssymb}
\usepackage{times}
\usepackage{textcomp}
\usepackage{verbatim}

\newcommand{\ptosfieldbis}{88.14\,V/cm}
\newcommand{\ptosfield}{79.94\,V/cm}
\newcommand{\stodfield}{80.42\,V/cm}
\newcommand{\ptodfield}{79.99\,V/cm}

\newcommand{\ptos}{$p\to s$}
\newcommand{\stod}{$s\to d$}
\newcommand{\ptod}{$p\to d$}

\begin{document}
\date{\today}

\title{Observation of a resonant four-body interaction in cold cesium
  Rydberg atoms}

\author{J.~H.~Gurian}\affiliation{Laboratoire Aim\'e Cotton, CNRS, Univ Paris-Sud, B\^at. 505, 91405 Orsay, France}
\author{P.~Cheinet}\affiliation{Laboratoire Aim\'e Cotton, CNRS, Univ Paris-Sud, B\^at. 505, 91405 Orsay, France}
\author{P.~Huillery}\affiliation{Laboratoire Aim\'e Cotton, CNRS, Univ Paris-Sud, B\^at. 505, 91405 Orsay, France}
\author{A.~Fioretti}\affiliation{Laboratoire Aim\'e Cotton, CNRS, Univ Paris-Sud, B\^at. 505, 91405 Orsay, France}
\author{J.~Zhao}\affiliation{Laboratoire Aim\'e Cotton, CNRS, Univ Paris-Sud, B\^at. 505, 91405 Orsay, France}\affiliation{State Key Laboratory
  of Quantum Optics and Quantum Optics Devices, College of Physics and
  Electronics Engineering, Shanxi University, Taiyuan 030006, China}
\author{P.L.~Gould}\affiliation{Laboratoire Aim\'e Cotton, CNRS, Univ Paris-Sud, B\^at. 505, 91405 Orsay, France}\affiliation{Department of
  Physics, University of Connecticut, Storrs, CT 06269-3046, U.S.A.}
\author{D.~Comparat}\affiliation{Laboratoire Aim\'e Cotton, CNRS, Univ
  Paris-Sud, B\^at. 505, 91405 Orsay, France} 
\author{P.~Pillet}\affiliation{Laboratoire Aim\'e Cotton, CNRS, Univ Paris-Sud, B\^at. 505, 91405 Orsay, France}

\begin{abstract}
  Cold Rydberg atoms subject to long-range dipole-dipole interactions
  represent a particularly interesting system for exploring few-body
  interactions and probing the transition from 2-body physics to the
  many-body regime. In this work we report the direct observation of a
  resonant 4-body Rydberg interaction.  We exploit the occurrence of
  an accidental quasi-coincidence of a 2-body and a 4-body resonant
  Stark-tuned F\"orster process in cesium to observe a resonant energy
  transfer requiring the simultaneous interaction of at least four
  neighboring atoms.  These results are relevant for the
  implementation of quantum gates with Rydberg atoms and for further
  studies of many-body physics.
\end{abstract}
\maketitle
The physics of atomic systems at low densities ($n\leq 10^{13}$\,cm$^{-3}$) can generally be described in terms of the action
of electromagnetic fields and binary (2-body) interactions.  However,
a number of interesting effects arise when few-body or many-body
interactions come into play. Notable examples include: 3-body
recombination~\cite{PhysRevLett.91.240402, PhysRevLett.91.123201,
  PhysRevLett.83.1751}, leading to molecule formation in optical or
magnetic traps; trimer photoassociation~\cite{PhysRevLett.105.163201};
and Efimov physics, leading to trimers and more recently tetramer
formation~\cite{NatPhys.5.417, PhysRevLett.102.140401,
  PhysRevLett.103.163202, Science.326.1683}.

Cold, highly-excited (Rydberg) atoms~\cite{Gallagher1994,
  review_gall_pillet,JOSAB.27.00A208} are a promising playground
for many-body interactions, due to their strong and long-range
interactions together with the long interaction times available in a
cold sample. This was first revealed in
studies~\cite{mourachko1998,anderson1998} on the broadening of Rydberg
energy transfer resonances.  Subsequent work further explored the
broadening mechanisms~\cite{2004PhysRevA.70.031401,tanner:043002,
  2008PhRvL.100l3007R,JOSAB.27.00A208, PhysRevA.82.052501,
  tanner:043002, PhysRevA.73.032725}, the influence of the system
dimensionality~\cite{PhysRevA.73.032725}, and
dephasing~\cite{PhysRevA.65.063404}.  A renewed interest in cold
Rydberg systems has recently been triggered by their possible use in
quantum computation~\cite{RevModPhys.82.2313}. Long-range van der
Waals~\cite{2004PhRvL..93f3001T} or
dipole-dipole~\cite{2006PhRvL..97h3003V, 2007PhRvL..99g3002V}
interactions allow control of the atomic excitation in a given volume,
the so-called blockade effect, enabling the implementation of quantum
gates~\cite{2009NatPh...5..115G, 2010PhRvL.104a0502W}. In this
respect, a careful control over the number of contributing partners is
necessary as 3-body and 4-body effects could be large enough to
interfere with the computation process~\cite{PhysRevLett.92.077903}.

Another key feature of cold Rydberg atoms is the ability to tune
interactions by simply using an external electric field. In the
process known as a F\"orster resonance, the energy of the final
many-body state can be Stark tuned into resonance with the initial
state, leading to a resonant energy
transfer (FRET)~\cite{PhysRevLett.93.233001, 2010PhRvL.104g3003R}.  In this
case, modeling the system requires including multiple atoms and
solving
 the full three-dimensional many-body wave
equation~\cite{PhysRevA.73.032725,2008PhRvA..77c2723W}. However, the
number of atoms which must be included for accurate results is still
unclear~\cite{2007NuPhA.790..728R, 2009PhRvA..79d3420Y,
PhysRevA.80.052712}.  It is therefore important to study cases
where a small number of atoms play a dominant role.  This is the case
in the striking experiments on 3- or 4-body
recombination~\cite{PhysRevLett.102.140401, PhysRevLett.103.163202,
  Science.326.1683}, where the few-body process is characterized by a
resonant loss mechanism.
 
\begin{figure}
  \includegraphics[width=0.48\textwidth]{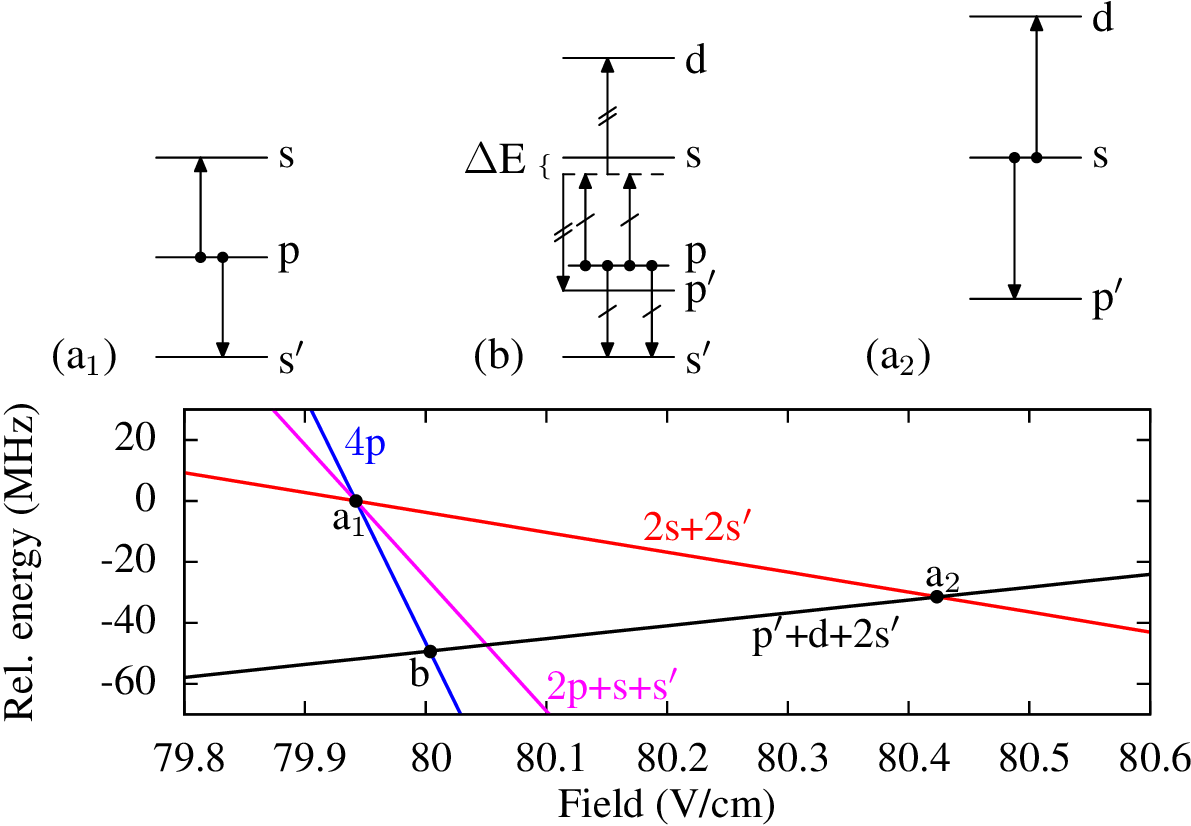}
  \caption{(Color online) Calculated four-atom energy for the two
    relevant 2-body resonances (a$_1$) and (a$_2$) and the \ptod~4-body
    resonance (b) as a function of the electric field. }
  \label{fig:energydiff}
\end{figure}

In this Letter we present first results on a 4-body energy transfer
process due to a 4-body Stark-tuned FRET resonance occurring
in cesium at an electric field of \ptodfield:
\begin{eqnarray}
  4\times 23p_{3/2}&\rightarrow& 2\times23s+23p_{1/2}+23d_{5/2}\label{eq:fourbody}.
\end{eqnarray}
We take advantage of two nearly resonant Cs 2-body Stark-tuned
FRET resonances,
\begin{eqnarray}
  23p_{3/2} + 23p_{3/2} &\rightarrow& 23s+ 24s \label{eq:ptos}\\
  24s + 24s &\rightarrow& 23p_{1/2}+ 23d_{5/2}\label{eq:stod},
\end{eqnarray}
occurring respectively at \ptosfield~and \stodfield. All states have
$|m_J|=1/2$ unless specified otherwise.  For convenience, the states
$23p_{3/2}$, $23s$, $24s$, $23p_{1/2}$ and $23d_{5/2}$ are labeled as
$p$, $s'$, $s$, $p'$, and $d$, respectively.  The three resonances are
illustrated in Fig.~\ref{fig:energydiff} and are denoted as \ptod~
(Eq.~\eqref{eq:fourbody}), \ptos~(Eq.~\eqref{eq:ptos}) and \stod~(Eq.~\eqref{eq:stod}).  Observing $d$ population after exciting the
$p$ state is the signature that four atoms exchanged energy.

The experimental setup consists of a standard Cs magneto-optical trap
(MOT) at the center of four parallel 60\,mm by 130\,mm wire mesh grids
of 80\,$\mu$m thickness and 1\,mm grid spacing.  The center pair of
grids is spaced by $1.845\pm0.01$\,cm, and the outer grids are 1.5\,cm
from the inner grids.  Voltages up to $\pm5$\,kV can be applied
arbitrarily to the four grids. Six additional small electrodes
surround the excitation region at the grid edges to cancel stray
fields.  The central grid spacing has been calibrated by
measuring Stark shifts of the $23p$ and $22d$ states for fields up to
100\,V/cm and comparing them to theory~\cite{PhysRevA.20.2251}.

The trapped atoms are excited to a Stark-shifted Rydberg state, $nl$,
using the MOT light and two additional lasers via the scheme: $6s \to
6p \to 7s \to nl$. The $6p \to 7s$ step uses a 1470\,nm, 10\,mW laser
frequency-locked on a Doppler-free feature in a Cs cell excited by
resonant $6s\to 6p$ light. In order to avoid perturbations of the
atomic cloud, the 1470\,nm laser is switched on for only 500\,ns, at a
10\,Hz repetition rate, by an acousto-optical modulator (AOM). The
first order AOM output is focused to a 400\,$\mu$m 
diameter spot at the atomic sample and its intensity is chosen to avoid
power-broadening the transition. A cw Ti:sapphire ring laser,
providing roughly 1.8\,W at $\approx795$\,nm, drives the $7s \to nl$
transition. This laser is switched in the same manner and the beam is
focused to a 300\,$\mu$m spot diameter and perpendicularly overlapped
with the 1470\,nm beam in the atomic sample. The simultaneous 500\,ns
pulses are short enough to avoid excitation blockade. We thus excite a
Gaussian cloud of up to $2\times 10^5$ $p$ atoms, 260\,$\mu$m in
diameter, with a maximum peak density, $\rho_p$, of $9\times
10^9$\,cm$^{-3}$.

\begin{figure}
  \includegraphics[width=0.48\textwidth]{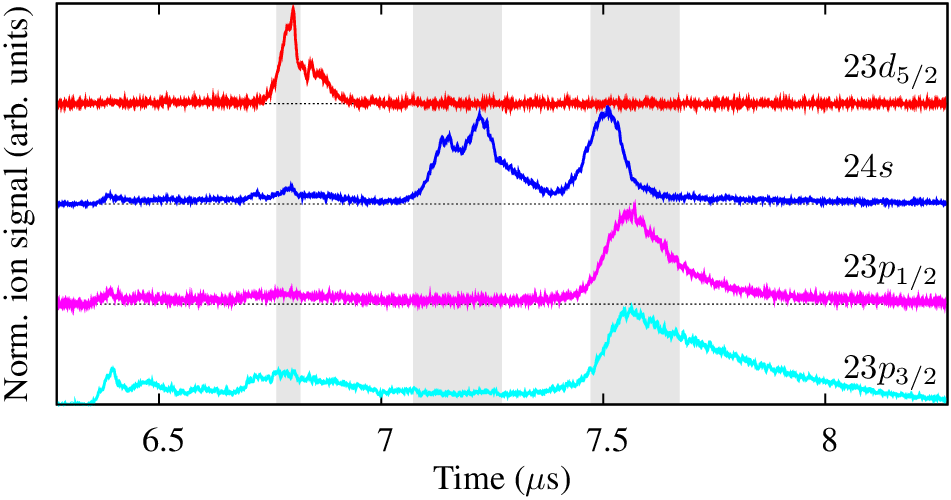}
  \caption{(Color online) Thirty-shot-averaged normalized time-of-flight signals
    for laser excitation to the Stark-mixed $d$, $s$, $p'$ and $p$
    states.  The four traces have been vertically offset for
    clarity, with the respective zero levels shown as horizontal
    dashed lines. The electric field is set to 79\,V/cm, far from any
    FRET resonance.  The gates for the $d$, $s$, and $p$ signals
    are shown as gray shaded regions. 
  }
  \label{fig:traces1}
\end{figure}

Selective field ionization is used to measure the populations of the
various Rydberg states.  A voltage ramp is applied to one of the grids,
$1.5\,\mu$s after the beginning of the laser excitation, rising to
4.3\,kV in 4\,$\mu$s. The resultant ions are detected by a
microchannel plate (MCP) detector, 210\,mm from the center of the
trapped cloud.  The amplitude of the field ionization pulse is chosen
to optimally isolate the $d$ time-of-flight (TOF) from the
other signals, as displayed in Fig.~\ref{fig:traces1}, where each
state has been excited directly in a non-resonant electric field. We
use the time gates shown to measure the population of each state.

It is important in this experiment to retrieve the correct populations
of the $p$, $s$, and $d$ states from a single TOF
signal. As one can see in Fig.~\ref{fig:traces1}, we have the ability
to differentiate these Rydberg states, but it is also apparent that
they are not cleanly separated into their respective gates, for
example due to blackbody radiation or multiple ionization pathways. We
therefore use traces, shown in Fig.~\ref{fig:traces1},
where each state is separately excited and the electric field is not
resonant, to quantify the crosstalk between the various gates. We also
account for the detection efficiency of each state, using a voltage
which ionizes completely, and separately measuring the global MCP
detection efficiency. We notice a slight
overestimation of the measured cross-talk coefficients leading to
negative $s$ or $d$ population for small $p$ excitation.  We correct
these coefficients and these corrections are included in the error bars shown in the
experimental figures below.
Throughout the paper, the recorded signals are
converted to the actual number of atoms in each state via the matrix,
\begin{equation}
  \label{eq:1}
  \begin{pmatrix}
    d\\ s\\ p
  \end{pmatrix}=
  \begin{pmatrix}
    2.016&-0.0644&-0.082\\
    -0.100&4.645&-0.275\\
    0.083&-3.147&4.149
  \end{pmatrix}
  \begin{pmatrix}
    d\\s\\p
  \end{pmatrix}_{\rm gate},
\end{equation}
whose off-diagonal elements characterize the crosstalk. Fortunately,
the $d$-state signal has minimal crosstalk with the
others, and a more advanced signal analysis~\cite{PhysRevA.52.3809} is not required.

\begin{figure}
  \includegraphics[width=0.48\textwidth]{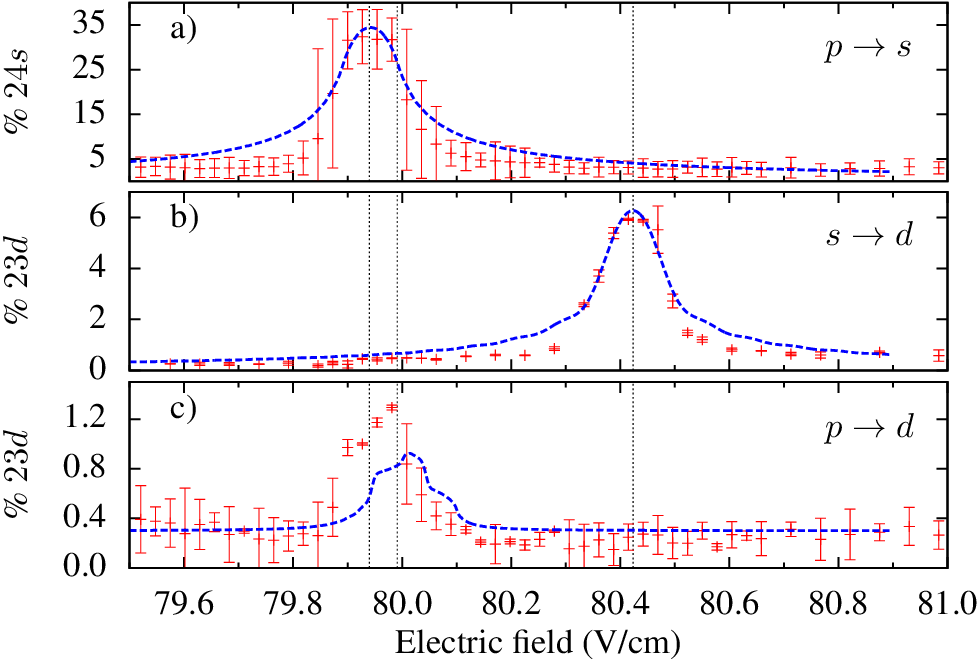}
  \caption{(Color online) Number of detected Rydberg atoms as a
    function of applied electric field.  a) percentage of detected $s$
    atoms when exciting the $p$ state.  b) percentage of $d$ detected
    when exciting the $s$ state.  c) percentage of $d$ detected when
    exciting the $p$ state.
    The error bars account for the corrections to the cross-talk
    coefficients, which are larger for $s$ than for $d$, and for the
    observed field inhomogeneity that affects more the $p$
    excitation, leading to larger error bars than in the $s$
    excitation case.  The results of our calculation for four
    equidistant atoms are overlaid as dashed blue lines, with the
    \ptos~calculation amplified by a factor of two
and the
    \stod~divided by a factor of three 
to coincide with
    experimental results.  The three resonant field values are
    illustrated by the vertical dotted lines.}
  \label{fig:transfer_multiplot}
\end{figure}

Although 2-body Stark-tuned FRET resonances have been observed
previously~\cite{2006PhRvL..97h3003V,mourachko1998}, we present them
below for completeness.  We first verify the 2-body \ptos~resonance,
as shown in Fig.~\ref{fig:transfer_multiplot}a.  The resonance is
observed at \ptosfield, and the flat-top resonance shape suggests that
the transition is saturated.  We observe a small field inhomogeneity
in our experimental region, which we estimate to be around 5 V/cm/cm,
via broadening of the $p$ laser excitation line.  This corresponds to
$\pm 0.05$\,V/cm in the excitation volume, which is thus the electric
field measurement resolution.  Stark field laser excitation of the $s$
state lets us measure the 2-body \stod~resonance, as shown in
Fig.~\ref{fig:transfer_multiplot}b.  We can excite up to
$1.4\times10^5$ $24s$ atoms at a peak density, $\rho_s$, of
$8\times10^9$\,cm$^{-3}$.  The \stod~resonance is observed at
\stodfield.  We have used the Cs energy levels
published by Sansonetti~\cite{JPhysChemRef38.761} for our calculations.  Unfortunately, the
$24p_{3/2}$ state energy uncertainty (0.03\,cm$^{-1}$) creates a large
uncertainty in the position of the 2-body resonances; $\pm 0.1$\,V/cm
for the \ptos~transition, and $\pm 1$\,V/cm for the
\stod~transition.  While the positions of the resonances are not well
defined, the slopes shown in Fig.~\ref{fig:energydiff} are stable to
within 0.1\%, and we have therefore aligned the calculated 2-body
resonances with our experimentally measured positions.

Once the 2-body resonances have been measured, the location of the
4-body resonance is accurately known.  
We observe greater than 1\% \ptod~transfer with a
peak at \ptodfield, as shown in Fig.~\ref{fig:transfer_multiplot}c.
While the 4-body resonance partially overlaps the \ptos~2-body
resonance in field, it is important to recall that the $d$ state
signal is well separated in the TOF.

\begin{figure}
  \centering
  \includegraphics[width=0.48\textwidth]{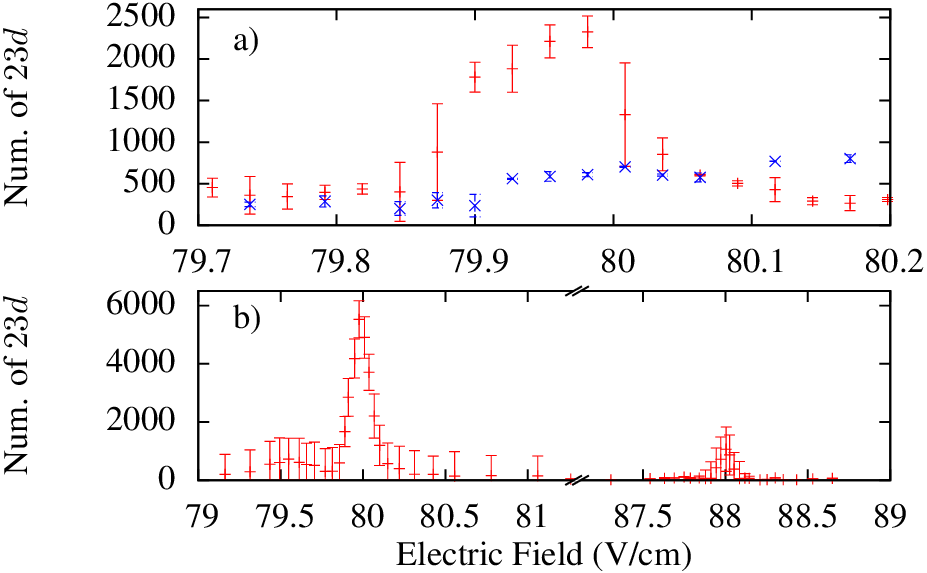}
  \caption{(Color online) Figure a) shows the number of detected $d$
    atoms as a function of the applied electric field for exciting the
    $p$ (red $+$) and $s$ (blue $\times$) states, with an $s$ density
    comparable or greater in the $s$ excitation than in the $p$
    excitation.  The on-resonant 4-body process creates more than a
    factor of four more $d$ atoms than the off-resonant \stod~2-body
    process at \ptodfield.  Figure b) shows the number of detected $d$
    atoms when exciting the \ptos~resonance for the $|m_J|=1/2$
    (\ptosfield) and $|m_J|=3/2$ (\ptosfieldbis).  We observe no
    significant $d$ state population in the $|m_J|=3/2$ case.  }
  \label{fig:numbercomparison}
\end{figure}

To provide insight into the 4-body resonance, we have developed a
minimal toy model with four equidistant atoms arranged as a
tetrahedron. The four possible states $|pppp\rangle$ (initial state),
$|ss^\prime pp\rangle$, $|ss^\prime ss^\prime\rangle$, and $|ds^\prime
p^\prime s^\prime \rangle$ (detected
state) are coupled by dipole-dipole interactions, calculated between
the in-field eigenstates of the Rydberg
atoms~\cite{PhysRevA.20.2251}. The final populations, shown in
Fig.~\ref{fig:transfer_multiplot} as the blue dashed curves, are
calculated using the density matrix and the experimental peak density
and field inhomogeneity. We average the results assuming an Erlang
(nearest neighbor) distribution for the 2-body distance between
the atoms, and a cubic Erlang distribution for the 4-body case.  Such
a 2-body model is not expected to precisely match the
experiment~\cite{PhysRevA.73.032725} and the calculated 2-body
\ptos~curve is amplified by a factor of two while the \stod~curve
has been diminished by a factor of three to match the
experimentally observed results.  To account for the 0.3\% background
observed in the \ptod~4-body transfer, the 4-body curve baseline has
been shifted accordingly.  While a more detailed many-body calculation
would be needed to reproduce the data~\cite{2008PhRvA..77c2723W}, it
is remarkable that such a crude 4-body calculation qualitatively
reproduces the shape of the experimental signal.

The observation of $d$ state population constitutes a clear,
direct signature of an interaction involving at least four bodies.
The strong signature that the process is not a simple combination of
two consecutive 2-body processes, but a genuine 4-body process, lies
in the relative strengths of the $d$ transfer at the 4-body resonance
field when initially exciting the $p$ or $s$
state. Figure~\ref{fig:numbercomparison}a shows the number of detected
$d$ atoms for comparable densities of $s$, either excited directly
($1.4\times10^5$) or obtained from exciting $p$ and allowing 2-body
transfer into $s$ ($<10^5$).  The larger number of detected $d$ atoms
(about a factor of four) when exciting $p$, despite a smaller $s$
density, is explained by the fact that here the 4-body \ptod~transfer
is resonant whereas the 2-body \stod~transfer is not.  Finally,
exciting the $p$ state in the $|m_J|=3/2$ case, where the
\ptos~resonance lies around 88.1\,V/cm and the 4-body resonance is
well separated from both 2-body resonances, we see in
Fig.~\ref{fig:numbercomparison}b that no significant $d$ population is
detected at the \ptos~resonance. The small observed signal is
compatible with the estimated error on the inversion matrix
coefficients in Eq.~\eqref{eq:1}.

A way to increase the population transfer is to shift the applied
field from the \ptos~resonance to the \stod~resonance
between laser excitation and detection. The two 2-body FRETs are then
consecutively resonant and we should get a $d$ transfer comparable to
that obtained when exciting $s$ directly. Indeed, we have observed up
to 7.5\% \ptod~population transfer with a 0.6\,V/cm field shift,
confirming that the \ptos~population transfer at \ptosfield~leads to
about the same $s$ density as direct $s$ excitation. With a larger
shift it is also possible to induce a transfer to $d$ from the
$|m_J|=3/2$ $p$ state. We observed up to 2.1\% \ptod~population
transfer with a -7.7\,V/cm field shift starting from \ptosfieldbis.

\begin{figure}
  \includegraphics[width=0.48\textwidth]{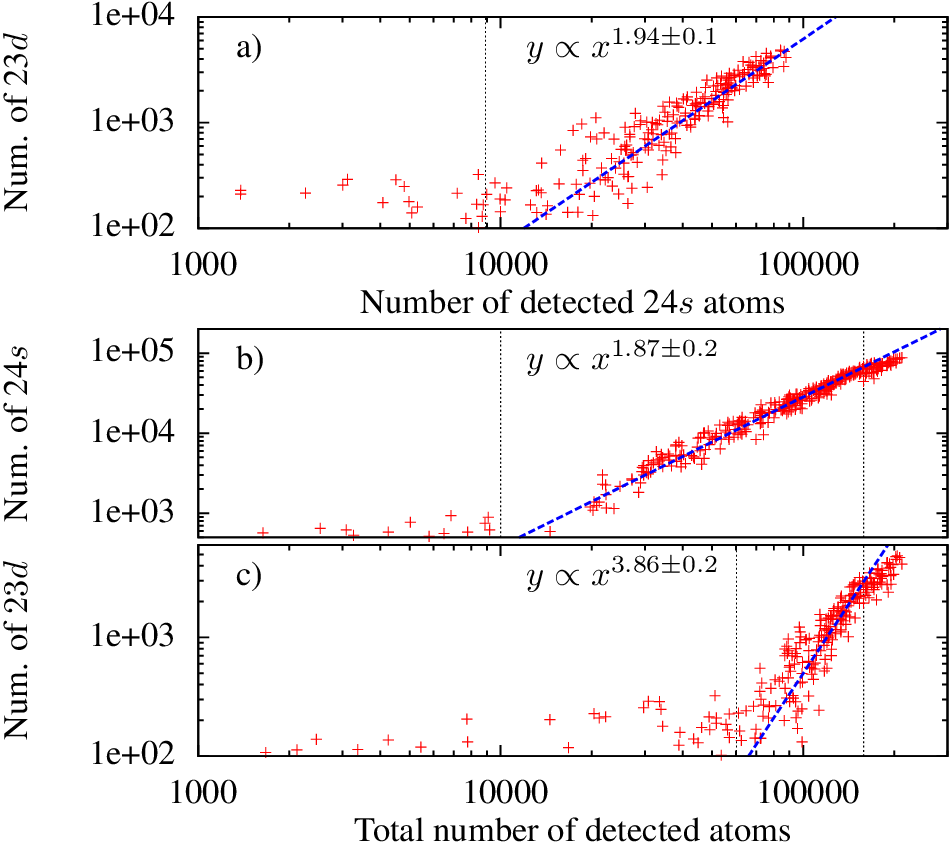}
  \caption{(Color online) Number of detected Rydberg atoms as the
    intensity of the Rydberg excitation laser is varied.  a) Number of
    $d$ atoms as a function of $s$ detected atoms.  b)
    Number of detected $s$ atoms as a function of the total number
    of detected Rydberg atoms.  c) Number of $d$ atoms created
    vs the total number of detected Rydberg atoms. Best fits below
    $1.6\times10^5$ atoms (demarcated with a vertical dotted line) are
    shown as the dashed blue lines.}
  \label{fig:intensity_multiplot}
\end{figure}

The field inhomogeneity and the proximity of the 4-body and 2-body
FRET resonances impede further studies on the resonance shape.
Nevertheless, having identified the 4-body resonance, we can study the
transfer dependence on the initial $p$ Rydberg atom density. We vary
this density by attenuating the Rydberg excitation laser with a set of
neutral density filters, sitting on the \ptodfield\,\ptod\,resonance.
The results are shown in Fig.~\ref{fig:intensity_multiplot}. We have
fit data only above the detection sensitivity of 500 atoms for $s$ and
100 atoms for $d$.
The population transfer from the initial $p$ state to
the $s$ state is shown in Fig.~\ref{fig:intensity_multiplot}b.  It
increases as $\rho_p^2$ until the
\ptos\,transfer starts saturating above $1.6\times10^5$ atoms, as expected from
Fig.~\ref{fig:transfer_multiplot},  and we
have limited all three fits to below this value.  The $d$ state population
transfer as a function of the detected $s$
population is shown in Fig.~\ref{fig:intensity_multiplot}a.  The
quadratic, i.e. nonlinear, dependence ensures us of the correct data
treatment presented in Eq.~\eqref{eq:1}.  Finally, we observe in
Fig.~\ref{fig:intensity_multiplot}c the number of detected $d$ atoms
as a function of the total number of excited Rydberg atoms.
We clearly see the influence of the $s$ transfer saturation on the $d$
transfer. Nevertheless, we can see the $d$
population transfer scaling as $\rho_p^4$, again demonstrating that
$d$ state observation links to a 4-body process.

In conclusion, we have observed a 4-body interaction
between Rydberg atoms excited in a MOT. The observed process is a
FRET involving the simultaneous
interaction of at least four neighboring partners. Its occurrence
depends on the electric field proximity of the parent 2-body
FRET resonances and to the quasi-coincidence of the 4-body
resonance with one of them.  We have studied the 4-body reaction as
a function of the electric field and the initial Rydberg density by
selectively detecting the reaction products.

Tunable FRET resonances are extremely interesting tools for
making Rydberg atoms into suitable systems for quantum information
storage and
processing~\cite{RevModPhys.82.2313,JOSAB.27.00A208}. Nevertheless,
care should be taken to properly evaluate the occurrence of
higher-order (3-body, 4-body or n-body) processes when issues of
fidelity and decoherence are to be addressed.

This work on a strong 4-body interaction is an important step towards
studying few-body and many-body effects in dilute gases.  Such
processes which go beyond the 2-body interaction are likely to occur
for many Rydberg states, especially at the larger densities that can
be obtained in optical traps. For instance, other 4-body FRET
resonances exist for four non-identical Rydberg atoms, for which the
resonance field is better separated from the 2-body resonance field.
Furthermore, the use of optical lattices to induce spatial order in
the system should allow new insights into the physics of 
novel quantum systems.

This work has been supported by the {\it Institut francilien de
  recherche sur les atomes froids} (IFRAF).  J.~G. and A.~F.  have been
supported by the ``\emph{Triangle de la Physique}'' under contracts
2007-n.74T and 2009-035T ``GULFSTREAM''.  P.~G. has been supported by
the ``\emph{Triangle de la Physique}'' under contract 2010-026T ``COCORYCO''.
 J.~Z. has been supported by the NSFC under grant No. 61078001.  

\bibliographystyle{apsrev4-1}

\begin{thebibliography}{10}%
\makeatletter
\providecommand \@ifxundefined [1]{%
 \ifx #1\undefined \expandafter \@firstoftwo
 \else \expandafter \@secondoftwo
\fi
}%
\providecommand \@ifnum [1]{%
 \ifnum #1\expandafter \@firstoftwo
 \else \expandafter \@secondoftwo
\fi
}%
\providecommand \enquote [1]{``#1''}%
\providecommand \bibnamefont  [1]{#1}%
\providecommand \bibfnamefont [1]{#1}%
\providecommand \citenamefont [1]{#1}%
\providecommand\href[0]{\@sanitize\@href}%
\providecommand\@href[1]{\endgroup\@@startlink{#1}\endgroup\@@href}%
\providecommand\@@href[1]{#1\@@endlink}%
\providecommand \@sanitize [0]{\begingroup\catcode`\&12\catcode`\#12\relax}%
\@ifxundefined \pdfoutput {\@firstoftwo}{%
 \@ifnum{\z@=\pdfoutput}{\@firstoftwo}{\@secondoftwo}%
}{%
 \providecommand\@@startlink[1]{\leavevmode}%
 \providecommand\@@endlink[0]{}%
}{%
 \providecommand\@@startlink[1]{%
  \leavevmode
  \pdfstartlink
   attr{/Border[0 0 1 ]/H/I/C[0 1 1]}%
   user{/Subtype/Link/A<</Type/Action/S/URI/URI(#1)>>}%
  \relax
 }%
 \providecommand\@@endlink[0]{\pdfendlink}%
}%
\providecommand \url  [0]{\begingroup\@sanitize \@url }%
\providecommand \@url [1]{\endgroup\@href {#1}{\urlprefix}}%
\providecommand \urlprefix [0]{URL }%
\providecommand \Eprint[0]{\href }%
\@ifxundefined \urlstyle {%
  \providecommand \doi [1]{doi:\discretionary{}{}{}#1}%
}{%
  \providecommand \doi [0]{doi:\discretionary{}{}{}\begingroup
  \urlstyle{rm}\Url }%
}%
\providecommand \doibase [0]{http://dx.doi.org/}%
\providecommand \Doi[1]{\href{\doibase#1}}%
\providecommand \bibAnnote [3]{%
  \BibitemShut{#1}%
  \begin{quotation}\noindent
    \textsc{Key:}\ #2\\\textsc{Annotation:}\ #3%
  \end{quotation}%
}%
\providecommand \bibAnnoteFile [2]{%
  \IfFileExists{#2}{\bibAnnote {#1} {#2} {\input{#2}}}{}%
}%
\providecommand \typeout [0]{\immediate \write \m@ne }%
\providecommand \selectlanguage [0]{\@gobble}%
\providecommand \bibinfo [0]{\@secondoftwo}%
\providecommand \bibfield [0]{\@secondoftwo}%
\providecommand \translation [1]{[#1]}%
\providecommand \BibitemOpen[0]{}%
\providecommand \bibitemStop [0]{}%
\providecommand \bibitemNoStop [0]{.\EOS\space}%
\providecommand \EOS [0]{\spacefactor3000\relax}%
\providecommand \BibitemShut [1]{\csname bibitem#1\endcsname}%
\bibitem{PhysRevLett.91.240402}%
  \BibitemOpen
  \bibfield{author}{%
  \bibinfo {author} {\bibfnamefont{S.}~\bibnamefont{Jochim}}, \bibinfo {author}
  {\bibfnamefont{M.}~\bibnamefont{Bartenstein}}, \bibinfo {author}
  {\bibfnamefont{A.}~\bibnamefont{Altmeyer}}, \bibinfo {author}
  {\bibfnamefont{G.}~\bibnamefont{Hendl}}, \bibinfo {author}
  {\bibfnamefont{C.}~\bibnamefont{Chin}}, \bibinfo {author}
  {\bibfnamefont{J.~H.}\ \bibnamefont{Denschlag}},\ and\ \bibinfo {author}
  {\bibfnamefont{R.}~\bibnamefont{Grimm}},\ }%
  \bibfield{journal}{%
  \Doi{10.1103/PhysRevLett.91.240402}{\bibinfo {journal} {Phys. Rev. Lett.}}\
  }%
  \textbf{\bibinfo {volume} {91}},\ \bibinfo {pages} {240402} (\bibinfo {month}
  {Dec}\ \bibinfo {year} {2003})%
  \bibAnnoteFile{NoStop}{PhysRevLett.91.240402}%
\bibitem{PhysRevLett.91.123201}%
  \BibitemOpen
  \bibfield{author}{%
  \bibinfo {author} {\bibfnamefont{T.}~\bibnamefont{Weber}}, \bibinfo {author}
  {\bibfnamefont{J.}~\bibnamefont{Herbig}}, \bibinfo {author}
  {\bibfnamefont{M.}~\bibnamefont{Mark}}, \bibinfo {author}
  {\bibfnamefont{H.-C.}\ \bibnamefont{N\"agerl}},\ and\ \bibinfo {author}
  {\bibfnamefont{R.}~\bibnamefont{Grimm}},\ }%
  \bibfield{journal}{%
  \Doi{10.1103/PhysRevLett.91.123201}{\bibinfo {journal} {Phys. Rev. Lett.}}\
  }%
  \textbf{\bibinfo {volume} {91}},\ \bibinfo {pages} {123201} (\bibinfo {month}
  {Sep}\ \bibinfo {year} {2003})%
  \bibAnnoteFile{NoStop}{PhysRevLett.91.123201}%
\bibitem{PhysRevLett.83.1751}%
  \BibitemOpen
  \bibfield{author}{%
  \bibinfo {author} {\bibfnamefont{B.~D.}\ \bibnamefont{Esry}}, \bibinfo
  {author} {\bibfnamefont{C.~H.}\ \bibnamefont{Greene}},\ and\ \bibinfo
  {author} {\bibfnamefont{J.~P.}\ \bibnamefont{Burke}},\ }%
  \bibfield{journal}{%
  \Doi{10.1103/PhysRevLett.83.1751}{\bibinfo {journal} {Phys. Rev. Lett.}}\ }%
  \textbf{\bibinfo {volume} {83}},\ \bibinfo {pages} {1751} (\bibinfo {month}
  {Aug}\ \bibinfo {year} {1999})%
  \bibAnnoteFile{NoStop}{PhysRevLett.83.1751}%
\bibitem{PhysRevLett.105.163201}%
  \BibitemOpen
  \bibfield{author}{%
  \bibinfo {author} {\bibfnamefont{V.}~\bibnamefont{Bendkowsky}}, \bibinfo
  {author} {\bibfnamefont{B.}~\bibnamefont{Butscher}}, \bibinfo {author}
  {\bibfnamefont{J.}~\bibnamefont{Nipper}}, \bibinfo {author}
  {\bibfnamefont{J.~B.}\ \bibnamefont{Balewski}}, \bibinfo {author}
  {\bibfnamefont{J.~P.}\ \bibnamefont{Shaffer}}, \bibinfo {author}
  {\bibfnamefont{R.}~\bibnamefont{L\"ow}}, \bibinfo {author}
  {\bibfnamefont{T.}~\bibnamefont{Pfau}}, \bibinfo {author}
  {\bibfnamefont{W.}~\bibnamefont{Li}}, \bibinfo {author}
  {\bibfnamefont{J.}~\bibnamefont{Stanojevic}}, \bibinfo {author}
  {\bibfnamefont{T.}~\bibnamefont{Pohl}},\ and\ \bibinfo {author}
  {\bibfnamefont{J.~M.}\ \bibnamefont{Rost}},\ }%
  \bibfield{journal}{%
  \Doi{10.1103/PhysRevLett.105.163201}{\bibinfo {journal} {Phys. Rev. Lett.}}\
  }%
  \textbf{\bibinfo {volume} {105}},\ \bibinfo {pages} {163201} (\bibinfo
  {month} {Oct}\ \bibinfo {year} {2010})%
  \bibAnnoteFile{NoStop}{PhysRevLett.105.163201}%
\bibitem{NatPhys.5.417}%
  \BibitemOpen
  \bibfield{author}{%
  \bibinfo {author} {\bibfnamefont{J.}~\bibnamefont{von Stecher}}, \bibinfo
  {author} {\bibfnamefont{J.}~\bibnamefont{D'Incao}},\ and\ \bibinfo {author}
  {\bibfnamefont{C.}~\bibnamefont{Greene}},\ }%
  \bibfield{journal}{%
  \bibinfo {journal} {Nat. Phys.}\ }%
  \textbf{\bibinfo {volume} {5}},\ \bibinfo {pages} {417} (\bibinfo {year}
  {2009})%
  \bibAnnoteFile{NoStop}{NatPhys.5.417}%
\bibitem{PhysRevLett.102.140401}%
  \BibitemOpen
  \bibfield{author}{%
  \bibinfo {author} {\bibfnamefont{F.}~\bibnamefont{Ferlaino}}, \bibinfo
  {author} {\bibfnamefont{S.}~\bibnamefont{Knoop}}, \bibinfo {author}
  {\bibfnamefont{M.}~\bibnamefont{Berninger}}, \bibinfo {author}
  {\bibfnamefont{W.}~\bibnamefont{Harm}}, \bibinfo {author}
  {\bibfnamefont{J.~P.}\ \bibnamefont{D'Incao}}, \bibinfo {author}
  {\bibfnamefont{H.-C.}\ \bibnamefont{N\"agerl}},\ and\ \bibinfo {author}
  {\bibfnamefont{R.}~\bibnamefont{Grimm}},\ }%
  \bibfield{journal}{%
  \Doi{10.1103/PhysRevLett.102.140401}{\bibinfo {journal} {Phys. Rev. Lett.}}\
  }%
  \textbf{\bibinfo {volume} {102}},\ \bibinfo {pages} {140401} (\bibinfo
  {month} {Apr}\ \bibinfo {year} {2009})%
  \bibAnnoteFile{NoStop}{PhysRevLett.102.140401}%
\bibitem{PhysRevLett.103.163202}%
  \BibitemOpen
  \bibfield{author}{%
  \bibinfo {author} {\bibfnamefont{N.}~\bibnamefont{Gross}}, \bibinfo {author}
  {\bibfnamefont{Z.}~\bibnamefont{Shotan}}, \bibinfo {author}
  {\bibfnamefont{S.}~\bibnamefont{Kokkelmans}},\ and\ \bibinfo {author}
  {\bibfnamefont{L.}~\bibnamefont{Khaykovich}},\ }%
  \bibfield{journal}{%
  \Doi{10.1103/PhysRevLett.103.163202}{\bibinfo {journal} {Phys. Rev. Lett.}}\
  }%
  \textbf{\bibinfo {volume} {103}},\ \bibinfo {pages} {163202} (\bibinfo
  {month} {Oct}\ \bibinfo {year} {2009})%
  \bibAnnoteFile{NoStop}{PhysRevLett.103.163202}%
\bibitem{Science.326.1683}%
  \BibitemOpen
  \bibfield{author}{%
  \bibinfo {author} {\bibfnamefont{S.~E.}\ \bibnamefont{Pollack}}, \bibinfo
  {author} {\bibfnamefont{D.}~\bibnamefont{Dries}},\ and\ \bibinfo {author}
  {\bibfnamefont{R.~G.}\ \bibnamefont{Hulet}},\ }%
  \bibfield{journal}{%
  \Doi{10.1126/science.1182840}{\bibinfo {journal} {Science}}\ }%
  \textbf{\bibinfo {volume} {326}},\ \bibinfo {pages} {1683} (\bibinfo {year}
  {2009})%
  \bibAnnoteFile{NoStop}{Science.326.1683}%
\bibitem{Gallagher1994}%
  \BibitemOpen
  \bibfield{author}{%
  \bibinfo {author} {\bibfnamefont{T.~F.}\ \bibnamefont{Gallagher}},\ }%
  \emph{\bibinfo {title} {Rydberg Atoms}}\ (\bibinfo {publisher} {Cambridge
  University Press},\ \bibinfo {address} {Cambridge},\ \bibinfo {year} {1994})%
  \bibAnnoteFile{NoStop}{Gallagher1994}%
\bibitem{review_gall_pillet}%
  \BibitemOpen
  \bibfield{author}{%
  \bibinfo {author} {\bibfnamefont{T.}~\bibnamefont{Gallagher}}\ and\ \bibinfo
  {author} {\bibfnamefont{P.}~\bibnamefont{Pillet}},\ }%
  \bibfield{journal}{%
  \bibinfo {journal} {Adv. At. Mol. Opt. Phys.}\ }%
  \textbf{\bibinfo {volume} {56}},\ \bibinfo {pages} {161} (\bibinfo {year}
  {2008})%
  \bibAnnoteFile{NoStop}{review_gall_pillet}%
\bibitem{JOSAB.27.00A208}%
  \BibitemOpen
  \bibfield{author}{%
  \bibinfo {author} {\bibfnamefont{D.}~\bibnamefont{Comparat}}\ and\ \bibinfo
  {author} {\bibfnamefont{P.}~\bibnamefont{Pillet}},\ }%
  \bibfield{journal}{%
  \Doi{10.1364/JOSAB.27.00A208}{\bibinfo {journal} {J. Opt. Soc. Am. B}}\ }%
  \textbf{\bibinfo {volume} {27}},\ \bibinfo {pages} {A208} (\bibinfo {month}
  {Jun}\ \bibinfo {year} {2010})%
  \bibAnnoteFile{NoStop}{JOSAB.27.00A208}%
\bibitem{mourachko1998}%
  \BibitemOpen
  \bibfield{author}{%
  \bibinfo {author} {\bibfnamefont{I.}~\bibnamefont{Mourachko}}, \bibinfo
  {author} {\bibfnamefont{D.}~\bibnamefont{Comparat}}, \bibinfo {author}
  {\bibfnamefont{F.}~\bibnamefont{de~Tomasi}}, \bibinfo {author}
  {\bibfnamefont{A.}~\bibnamefont{Fioretti}}, \bibinfo {author}
  {\bibfnamefont{P.}~\bibnamefont{Nosbaum}}, \bibinfo {author}
  {\bibfnamefont{V.~M.}\ \bibnamefont{Akulin}},\ and\ \bibinfo {author}
  {\bibfnamefont{P.}~\bibnamefont{Pillet}},\ }%
  \bibfield{journal}{%
  \bibinfo {journal} {Phys. Rev. Lett.}\ }%
  \textbf{\bibinfo {volume} {80}},\ \bibinfo {pages} {253} (\bibinfo {year}
  {1998})%
  \bibAnnoteFile{NoStop}{mourachko1998}%
\bibitem{anderson1998}%
  \BibitemOpen
  \bibfield{author}{%
  \bibinfo {author} {\bibfnamefont{W.~R.}\ \bibnamefont{Anderson}}, \bibinfo
  {author} {\bibfnamefont{J.~R.}\ \bibnamefont{Veale}},\ and\ \bibinfo {author}
  {\bibfnamefont{T.~F.}\ \bibnamefont{Gallagher}},\ }%
  \bibfield{journal}{%
  \bibinfo {journal} {Phys. Rev. Lett.}\ }%
  \textbf{\bibinfo {volume} {80}},\ \bibinfo {pages} {249} (\bibinfo {year}
  {1998})%
  \bibAnnoteFile{NoStop}{anderson1998}%
\bibitem{2004PhysRevA.70.031401}%
  \BibitemOpen
  \bibfield{author}{%
  \bibinfo {author} {\bibfnamefont{I.}~\bibnamefont{Mourachko}}, \bibinfo
  {author} {\bibfnamefont{W.}~\bibnamefont{Li}},\ and\ \bibinfo {author}
  {\bibfnamefont{T.~F.}\ \bibnamefont{Gallagher}},\ }%
  \bibfield{journal}{%
  \Doi{10.1103/PhysRevA.70.031401}{\bibinfo {journal} {Phys. Rev. A}}\ }%
  \textbf{\bibinfo {volume} {70}},\ \bibinfo {pages} {031401} (\bibinfo {month}
  {Sep}\ \bibinfo {year} {2004})%
  \bibAnnoteFile{NoStop}{2004PhysRevA.70.031401}%
\bibitem{tanner:043002}%
  \BibitemOpen
  \bibfield{author}{%
  \bibinfo {author} {\bibfnamefont{P.~J.}\ \bibnamefont{Tanner}}, \bibinfo
  {author} {\bibfnamefont{J.}~\bibnamefont{Han}}, \bibinfo {author}
  {\bibfnamefont{E.~S.}\ \bibnamefont{Shuman}},\ and\ \bibinfo {author}
  {\bibfnamefont{T.~F.}\ \bibnamefont{Gallagher}},\ }%
  \bibfield{journal}{%
  \Doi{10.1103/PhysRevLett.100.043002}{\bibinfo {journal} {Phys. Rev. Lett.}}\
  }%
  \textbf{\bibinfo {volume} {100}},\ \bibinfo {pages} {43002} (\bibinfo {year}
  {2008})%
  \bibAnnoteFile{NoStop}{tanner:043002}%
\bibitem{2008PhRvL.100l3007R}%
  \BibitemOpen
  \bibfield{author}{%
  \bibinfo {author} {\bibfnamefont{A.}~\bibnamefont{{Reinhard}}}, \bibinfo
  {author} {\bibfnamefont{T.}~\bibnamefont{{Cubel Liebisch}}}, \bibinfo
  {author} {\bibfnamefont{K.~C.}\ \bibnamefont{{Younge}}}, \bibinfo {author}
  {\bibfnamefont{P.~R.}\ \bibnamefont{{Berman}}},\ and\ \bibinfo {author}
  {\bibfnamefont{G.}~\bibnamefont{{Raithel}}},\ }%
  \bibfield{journal}{%
  \Doi{10.1103/PhysRevLett.100.123007}{\bibinfo {journal} {Phys. Rev. Lett.}}\
  }%
  \textbf{\bibinfo {volume} {100}},\ \bibinfo {pages} {123007} (\bibinfo
  {month} {Mar.}\ \bibinfo {year} {2008})%
  \bibAnnoteFile{NoStop}{2008PhRvL.100l3007R}%
\bibitem{PhysRevA.82.052501}%
  \BibitemOpen
  \bibfield{author}{%
  \bibinfo {author} {\bibfnamefont{J.}~\bibnamefont{Han}},\ }%
  \bibfield{journal}{%
  \Doi{10.1103/PhysRevA.82.052501}{\bibinfo {journal} {Phys. Rev. A}}\ }%
  \textbf{\bibinfo {volume} {82}},\ \bibinfo {pages} {052501} (\bibinfo {month}
  {Nov}\ \bibinfo {year} {2010})%
  \bibAnnoteFile{NoStop}{PhysRevA.82.052501}%
\bibitem{PhysRevA.73.032725}%
  \BibitemOpen
  \bibfield{author}{%
  \bibinfo {author} {\bibfnamefont{T.~J.}\ \bibnamefont{Carroll}}, \bibinfo
  {author} {\bibfnamefont{S.}~\bibnamefont{Sunder}},\ and\ \bibinfo {author}
  {\bibfnamefont{M.~W.}\ \bibnamefont{Noel}},\ }%
  \bibfield{journal}{%
  \Doi{10.1103/PhysRevA.73.032725}{\bibinfo {journal} {Phys. Rev. A}}\ }%
  \textbf{\bibinfo {volume} {73}},\ \bibinfo {pages} {032725} (\bibinfo {month}
  {Mar}\ \bibinfo {year} {2006})%
  \bibAnnoteFile{NoStop}{PhysRevA.73.032725}%
\bibitem{PhysRevA.65.063404}%
  \BibitemOpen
  \bibfield{author}{%
  \bibinfo {author} {\bibfnamefont{W.~R.}\ \bibnamefont{Anderson}}, \bibinfo
  {author} {\bibfnamefont{M.~P.}\ \bibnamefont{Robinson}}, \bibinfo {author}
  {\bibfnamefont{J.~D.~D.}\ \bibnamefont{Martin}},\ and\ \bibinfo {author}
  {\bibfnamefont{T.~F.}\ \bibnamefont{Gallagher}},\ }%
  \bibfield{journal}{%
  \Doi{10.1103/PhysRevA.65.063404}{\bibinfo {journal} {Phys. Rev. A}}\ }%
  \textbf{\bibinfo {volume} {65}},\ \bibinfo {pages} {063404} (\bibinfo {month}
  {Jun}\ \bibinfo {year} {2002})%
  \bibAnnoteFile{NoStop}{PhysRevA.65.063404}%
\bibitem{RevModPhys.82.2313}%
  \BibitemOpen
  \bibfield{author}{%
  \bibinfo {author} {\bibfnamefont{M.}~\bibnamefont{Saffman}}, \bibinfo
  {author} {\bibfnamefont{T.~G.}\ \bibnamefont{Walker}},\ and\ \bibinfo
  {author} {\bibfnamefont{K.}~\bibnamefont{M\"olmer}},\ }%
  \bibfield{journal}{%
  \Doi{10.1103/RevModPhys.82.2313}{\bibinfo {journal} {Rev. Mod. Phys.}}\ }%
  \textbf{\bibinfo {volume} {82}},\ \bibinfo {pages} {2313} (\bibinfo {month}
  {Aug}\ \bibinfo {year} {2010})%
  \bibAnnoteFile{NoStop}{RevModPhys.82.2313}%
\bibitem{2004PhRvL..93f3001T}%
  \BibitemOpen
  \bibfield{author}{%
  \bibinfo {author} {\bibfnamefont{D.}~\bibnamefont{{Tong}}}, \bibinfo {author}
  {\bibfnamefont{S.~M.}\ \bibnamefont{{Farooqi}}}, \bibinfo {author}
  {\bibfnamefont{J.}~\bibnamefont{{Stanojevic}}}, \bibinfo {author}
  {\bibfnamefont{S.}~\bibnamefont{{Krishnan}}}, \bibinfo {author}
  {\bibfnamefont{Y.~P.}\ \bibnamefont{{Zhang}}}, \bibinfo {author}
  {\bibfnamefont{R.}~\bibnamefont{{C{\^o}t{\'e}}}}, \bibinfo {author}
  {\bibfnamefont{E.~E.}\ \bibnamefont{{Eyler}}},\ and\ \bibinfo {author}
  {\bibfnamefont{P.~L.}\ \bibnamefont{{Gould}}},\ }%
  \bibfield{journal}{%
  \Doi{10.1103/PhysRevLett.93.063001}{\bibinfo {journal} {Phys. Rev. Lett.}}\
  }%
  \textbf{\bibinfo {volume} {93}},\ \bibinfo {pages} {063001} (\bibinfo {month}
  {Aug.}\ \bibinfo {year} {2004})%
  \bibAnnoteFile{NoStop}{2004PhRvL..93f3001T}%
\bibitem{2006PhRvL..97h3003V}%
  \BibitemOpen
  \bibfield{author}{%
  \bibinfo {author} {\bibfnamefont{T.}~\bibnamefont{{Vogt}}}, \bibinfo {author}
  {\bibfnamefont{M.}~\bibnamefont{{Viteau}}}, \bibinfo {author}
  {\bibfnamefont{J.}~\bibnamefont{{Zhao}}}, \bibinfo {author}
  {\bibfnamefont{A.}~\bibnamefont{{Chotia}}}, \bibinfo {author}
  {\bibfnamefont{D.}~\bibnamefont{{Comparat}}},\ and\ \bibinfo {author}
  {\bibfnamefont{P.}~\bibnamefont{{Pillet}}},\ }%
  \bibfield{journal}{%
  \Doi{10.1103/PhysRevLett.97.083003}{\bibinfo {journal} {Phys. Rev. Lett.}}\
  }%
  \textbf{\bibinfo {volume} {97}},\ \bibinfo {pages} {083003} (\bibinfo {month}
  {Aug.}\ \bibinfo {year} {2006})%
  \bibAnnoteFile{NoStop}{2006PhRvL..97h3003V}%
\bibitem{2007PhRvL..99g3002V}%
  \BibitemOpen
  \bibfield{author}{%
  \bibinfo {author} {\bibfnamefont{T.}~\bibnamefont{{Vogt}}}, \bibinfo {author}
  {\bibfnamefont{M.}~\bibnamefont{{Viteau}}}, \bibinfo {author}
  {\bibfnamefont{A.}~\bibnamefont{{Chotia}}}, \bibinfo {author}
  {\bibfnamefont{J.}~\bibnamefont{{Zhao}}}, \bibinfo {author}
  {\bibfnamefont{D.}~\bibnamefont{{Comparat}}},\ and\ \bibinfo {author}
  {\bibfnamefont{P.}~\bibnamefont{{Pillet}}},\ }%
  \bibfield{journal}{%
  \Doi{10.1103/PhysRevLett.99.073002}{\bibinfo {journal} {Phys. Rev. Lett.}}\
  }%
  \textbf{\bibinfo {volume} {99}},\ \bibinfo {pages} {073002} (\bibinfo {month}
  {Aug.}\ \bibinfo {year} {2007})%
  \bibAnnoteFile{NoStop}{2007PhRvL..99g3002V}%
\bibitem{2009NatPh...5..115G}%
  \BibitemOpen
  \bibfield{author}{%
  \bibinfo {author} {\bibfnamefont{A.}~\bibnamefont{{Ga{\"e}tan}}}, \bibinfo
  {author} {\bibfnamefont{Y.}~\bibnamefont{{Miroshnychenko}}}, \bibinfo
  {author} {\bibfnamefont{T.}~\bibnamefont{{Wilk}}}, \bibinfo {author}
  {\bibfnamefont{A.}~\bibnamefont{{Chotia}}}, \bibinfo {author}
  {\bibfnamefont{M.}~\bibnamefont{{Viteau}}}, \bibinfo {author}
  {\bibfnamefont{D.}~\bibnamefont{{Comparat}}}, \bibinfo {author}
  {\bibfnamefont{P.}~\bibnamefont{{Pillet}}}, \bibinfo {author}
  {\bibfnamefont{A.}~\bibnamefont{{Browaeys}}},\ and\ \bibinfo {author}
  {\bibfnamefont{P.}~\bibnamefont{{Grangier}}},\ }%
  \bibfield{journal}{%
  \Doi{10.1038/nphys1183}{\bibinfo {journal} {Nat. Phys.}}\ }%
  \textbf{\bibinfo {volume} {5}},\ \bibinfo {pages} {115} (\bibinfo {month}
  {Feb.}\ \bibinfo {year} {2009})%
  \bibAnnoteFile{NoStop}{2009NatPh...5..115G}%
\bibitem{2010PhRvL.104a0502W}%
  \BibitemOpen
  \bibfield{author}{%
  \bibinfo {author} {\bibfnamefont{T.}~\bibnamefont{{Wilk}}}, \bibinfo {author}
  {\bibfnamefont{A.}~\bibnamefont{{Ga{\"e}tan}}}, \bibinfo {author}
  {\bibfnamefont{C.}~\bibnamefont{{Evellin}}}, \bibinfo {author}
  {\bibfnamefont{J.}~\bibnamefont{{Wolters}}}, \bibinfo {author}
  {\bibfnamefont{Y.}~\bibnamefont{{Miroshnychenko}}}, \bibinfo {author}
  {\bibfnamefont{P.}~\bibnamefont{{Grangier}}},\ and\ \bibinfo {author}
  {\bibfnamefont{A.}~\bibnamefont{{Browaeys}}},\ }%
  \bibfield{journal}{%
  \Doi{10.1103/PhysRevLett.104.010502}{\bibinfo {journal} {Phys. Rev. Lett.}}\
  }%
  \textbf{\bibinfo {volume} {104}},\ \bibinfo {pages} {010502} (\bibinfo
  {month} {Jan.}\ \bibinfo {year} {2010})%
  \bibAnnoteFile{NoStop}{2010PhRvL.104a0502W}%
\bibitem{PhysRevLett.92.077903}%
  \BibitemOpen
  \bibfield{author}{%
  \bibinfo {author} {\bibfnamefont{A.}~\bibnamefont{Mizel}}\ and\ \bibinfo
  {author} {\bibfnamefont{D.~A.}\ \bibnamefont{Lidar}},\ }%
  \bibfield{journal}{%
  \Doi{10.1103/PhysRevLett.92.077903}{\bibinfo {journal} {Phys. Rev. Lett.}}\
  }%
  \textbf{\bibinfo {volume} {92}},\ \bibinfo {pages} {077903} (\bibinfo {month}
  {Feb}\ \bibinfo {year} {2004})%
  \bibAnnoteFile{NoStop}{PhysRevLett.92.077903}%
\bibitem{PhysRevLett.93.233001}%
  \BibitemOpen
  \bibfield{author}{%
  \bibinfo {author} {\bibfnamefont{K.}~\bibnamefont{Afrousheh}}, \bibinfo
  {author} {\bibfnamefont{P.}~\bibnamefont{Bohlouli-Zanjani}}, \bibinfo
  {author} {\bibfnamefont{D.}~\bibnamefont{Vagale}}, \bibinfo {author}
  {\bibfnamefont{A.}~\bibnamefont{Mugford}}, \bibinfo {author}
  {\bibfnamefont{M.}~\bibnamefont{Fedorov}},\ and\ \bibinfo {author}
  {\bibfnamefont{J.~D.~D.}\ \bibnamefont{Martin}},\ }%
  \bibfield{journal}{%
  \Doi{10.1103/PhysRevLett.93.233001}{\bibinfo {journal} {Phys. Rev. Lett.}}\
  }%
  \textbf{\bibinfo {volume} {93}},\ \bibinfo {pages} {233001} (\bibinfo {month}
  {Nov}\ \bibinfo {year} {2004})%
  \bibAnnoteFile{NoStop}{PhysRevLett.93.233001}%
\bibitem{2010PhRvL.104g3003R}%
  \BibitemOpen
  \bibfield{author}{%
  \bibinfo {author} {\bibfnamefont{I.~I.}\ \bibnamefont{{Ryabtsev}}}, \bibinfo
  {author} {\bibfnamefont{D.~B.}\ \bibnamefont{{Tretyakov}}}, \bibinfo {author}
  {\bibfnamefont{I.~I.}\ \bibnamefont{{Beterov}}},\ and\ \bibinfo {author}
  {\bibfnamefont{V.~M.}\ \bibnamefont{{Entin}}},\ }%
  \bibfield{journal}{%
  \Doi{10.1103/PhysRevLett.104.073003}{\bibinfo {journal} {Phys. Rev. Lett.}}\
  }%
  \textbf{\bibinfo {volume} {104}},\ \bibinfo {pages} {073003} (\bibinfo
  {month} {Feb.}\ \bibinfo {year} {2010})%
  \bibAnnoteFile{NoStop}{2010PhRvL.104g3003R}%
\bibitem{2008PhRvA..77c2723W}%
  \BibitemOpen
  \bibfield{author}{%
  \bibinfo {author} {\bibfnamefont{T.~G.}\ \bibnamefont{{Walker}}}\ and\
  \bibinfo {author} {\bibfnamefont{M.}~\bibnamefont{{Saffman}}},\ }%
  \bibfield{journal}{%
  \Doi{10.1103/PhysRevA.77.032723}{\bibinfo {journal} {Phys. Rev. A}}\ }%
  \textbf{\bibinfo {volume} {77}},\ \bibinfo {pages} {032723} (\bibinfo {month}
  {Mar.}\ \bibinfo {year} {2008})%
  \bibAnnoteFile{NoStop}{2008PhRvA..77c2723W}%
\bibitem{2007NuPhA.790..728R}%
  \BibitemOpen
  \bibfield{author}{%
  \bibinfo {author} {\bibfnamefont{M.}~\bibnamefont{{Reetz-Lamour}}}, \bibinfo
  {author} {\bibfnamefont{T.}~\bibnamefont{{Amthor}}}, \bibinfo {author}
  {\bibfnamefont{S.}~\bibnamefont{{Westermann}}}, \bibinfo {author}
  {\bibfnamefont{J.}~\bibnamefont{{Denskat}}}, \bibinfo {author}
  {\bibfnamefont{A.~L.}\ \bibnamefont{{de Oliveira}}},\ and\ \bibinfo {author}
  {\bibfnamefont{M.}~\bibnamefont{{Weidem{\"u}ller}}},\ }%
  \bibfield{journal}{%
  \Doi{10.1016/j.nuclphysa.2007.03.124}{\bibinfo {journal} {Nucl. Phys. A}}\ }%
  \textbf{\bibinfo {volume} {790}},\ \bibinfo {pages} {728} (\bibinfo {month}
  {Jun.}\ \bibinfo {year} {2007})%
  \bibAnnoteFile{NoStop}{2007NuPhA.790..728R}%
\bibitem{2009PhRvA..79d3420Y}%
  \BibitemOpen
  \bibfield{author}{%
  \bibinfo {author} {\bibfnamefont{K.~C.}\ \bibnamefont{{Younge}}}, \bibinfo
  {author} {\bibfnamefont{A.}~\bibnamefont{{Reinhard}}}, \bibinfo {author}
  {\bibfnamefont{T.}~\bibnamefont{{Pohl}}}, \bibinfo {author}
  {\bibfnamefont{P.~R.}\ \bibnamefont{{Berman}}},\ and\ \bibinfo {author}
  {\bibfnamefont{G.}~\bibnamefont{{Raithel}}},\ }%
  \bibfield{journal}{%
  \Doi{10.1103/PhysRevA.79.043420}{\bibinfo {journal} {Phys. Rev. A}}\ }%
  \textbf{\bibinfo {volume} {79}},\ \bibinfo {pages} {043420} (\bibinfo {month}
  {Apr.}\ \bibinfo {year} {2009})%
  \bibAnnoteFile{NoStop}{2009PhRvA..79d3420Y}%
\bibitem{PhysRevA.80.052712}%
  \BibitemOpen
  \bibfield{author}{%
  \bibinfo {author} {\bibfnamefont{T.~J.}\ \bibnamefont{Carroll}}, \bibinfo
  {author} {\bibfnamefont{C.}~\bibnamefont{Daniel}}, \bibinfo {author}
  {\bibfnamefont{L.}~\bibnamefont{Hoover}}, \bibinfo {author}
  {\bibfnamefont{T.}~\bibnamefont{Sidie}},\ and\ \bibinfo {author}
  {\bibfnamefont{M.~W.}\ \bibnamefont{Noel}},\ }%
  \bibfield{journal}{%
  \Doi{10.1103/PhysRevA.80.052712}{\bibinfo {journal} {Phys. Rev. A}}\ }%
  \textbf{\bibinfo {volume} {80}},\ \bibinfo {pages} {052712} (\bibinfo {month}
  {Nov}\ \bibinfo {year} {2009})%
  \bibAnnoteFile{NoStop}{PhysRevA.80.052712}%
\bibitem{PhysRevA.20.2251}%
  \BibitemOpen
  \bibfield{author}{%
  \bibinfo {author} {\bibfnamefont{M.~L.}\ \bibnamefont{Zimmerman}}, \bibinfo
  {author} {\bibfnamefont{M.~G.}\ \bibnamefont{Littman}}, \bibinfo {author}
  {\bibfnamefont{M.~M.}\ \bibnamefont{Kash}},\ and\ \bibinfo {author}
  {\bibfnamefont{D.}~\bibnamefont{Kleppner}},\ }%
  \bibfield{journal}{%
  \Doi{10.1103/PhysRevA.20.2251}{\bibinfo {journal} {Phys. Rev. A}}\ }%
  \textbf{\bibinfo {volume} {20}},\ \bibinfo {pages} {2251} (\bibinfo {month}
  {Dec}\ \bibinfo {year} {1979})%
  \bibAnnoteFile{NoStop}{PhysRevA.20.2251}%
\bibitem{PhysRevA.52.3809}%
  \BibitemOpen
  \bibfield{author}{%
  \bibinfo {author} {\bibfnamefont{V.~D.}\ \bibnamefont{Irby}}, \bibinfo
  {author} {\bibfnamefont{R.~G.}\ \bibnamefont{Rolfes}}, \bibinfo {author}
  {\bibfnamefont{O.~P.}\ \bibnamefont{Makarov}}, \bibinfo {author}
  {\bibfnamefont{K.~B.}\ \bibnamefont{MacAdam}},\ and\ \bibinfo {author}
  {\bibfnamefont{M.~I.}\ \bibnamefont{Syrkin}},\ }%
  \bibfield{journal}{%
  \Doi{10.1103/PhysRevA.52.3809}{\bibinfo {journal} {Phys. Rev. A}}\ }%
  \textbf{\bibinfo {volume} {52}},\ \bibinfo {pages} {3809} (\bibinfo {month}
  {Nov}\ \bibinfo {year} {1995})%
  \bibAnnoteFile{NoStop}{PhysRevA.52.3809}%
\bibitem{JPhysChemRef38.761}%
  \BibitemOpen
  \bibfield{author}{%
  \bibinfo {author} {\bibfnamefont{J.~E.}\ \bibnamefont{Sansonetti}},\ }%
  \bibfield{journal}{%
  \bibinfo {journal} {J. Phys. Chem. Ref. Data}\ }%
  \textbf{\bibinfo {volume} {38}},\ \bibinfo {pages} {761} (\bibinfo {year}
  {2009})%
  \bibAnnoteFile{NoStop}{JPhysChemRef38.761}%
\end{thebibliography}
\end{document}